%% file: main.tex
\documentclass{IEEEcsmag}

\usepackage{amsmath,amssymb,amsfonts}
\usepackage{algorithmic}
\usepackage{graphicx}
\usepackage{textcomp}
\usepackage{dcolumn}
\usepackage{bm}
\usepackage{tikz}
\usetikzlibrary{arrows.meta,positioning,calc, shapes.geometric}
\usepackage{mathtools}
\usepackage{tabularx}
\usepackage{physics}
\usepackage[normalem]{ulem}
\usepackage{braket}
\usepackage{multirow}
\usepackage{booktabs}
\usepackage{stackengine}
\usepackage{array}
\usepackage{xcolor}
\usepackage{lipsum}
\usepackage{dblfloatfix} 
\usetikzlibrary{arrows,shadows,positioning}
\usepackage[caption=false]{subfig}
\usepackage{lipsum}
\usepackage{array}
\usepackage{stmaryrd}
\usepackage{relsize}
\usepackage{float}

\definecolor{cincinnati-red}{RGB}{190,0,0}
\definecolor{ForestGreen}{RGB}{34,139,34}

\usepackage[colorlinks,urlcolor=blue,linkcolor=blue,citecolor=blue]{hyperref}
\expandafter\def\expandafter\UrlBreaks\expandafter{\UrlBreaks\do\/\do\*\do\-\do\~\do\'\do\"\do\-}
\usepackage{upmath,color}

\jvol{XX}
\jnum{XX}
\paper{8}
\jmonth{September}
\jname{Publication Name}
\jtitle{Quantum Repeater Protocol using Quantum
Error Correction for Distillation}
\pubyear{2025}

\newtheorem{lemma}{Lemma}

\begin{document}

\sptitle{Special Issue on Quantum Networks}

\title{Quantum Repeater Protocol using Quantum Error Correction for Distillation}
\author{Ashlesha Patil}
 \affil{Wyant College of Optical Sciences, University of Arizona, Tucson, AZ, USA.\\ NSF Center for Quantum Networks.}

 \author{Michele Pacenti}
 \affil{Department of Electrical and Computer Engineering, University of Arizona, Tucson, AZ, USA.\\  NSF Center for Quantum Networks.}

\author{Bane Vasi\'c}
 \affil{Department of Electrical and Computer Engineering, University of Arizona, Tucson, AZ, USA.\\  NSF Center for Quantum Networks.}

\author{Saikat Guha}
 \affil{Department of Electrical and Computer Engineering, University of Maryland, College Park, MD.\\  NSF Center for Quantum Networks.}

\author{Narayanan Rengaswamy}
 \affil{Department of Electrical and Computer Engineering, University of Arizona, Tucson, AZ, USA.\\  NSF Center for Quantum Networks.}

\markboth{Special Issue on Quantum Networks}{Special Issue on Quantum Networks}

\begin{abstract}
\looseness-1Bell-state measurement (BSM) on entangled states shared between quantum repeaters is the fundamental operation used to route entanglement in
quantum networks. Performing BSMs on Werner states shared between repeaters leads to exponential decay in the fidelity of the end-to-end Werner state with the
number of repeaters, necessitating entanglement distillation. In this work, we use quantum error correcting codes for deterministic entanglement distillation to route Werner states on a chain of repeaters. To maximize the end-to-end distillable entanglement, we utilize global link-state knowledge to determine the optimal policy for scheduling distillation and BSMs at the repeaters.
We observe that low-rate codes produce high-fidelity end-to-end states owing to their excellent error-correcting capability, whereas high-rate codes yield a larger number of end-to-end states but of lower fidelity. The number of quantum memories used at repeaters increases with the code rate as well as the classical computation time of the decoder.

\end{abstract}

\maketitle

\input{sections/introduction.tex}

\input{sections/background.tex}
\input{sections/protocol.tex}
\input{sections/results.tex}

\def\refname{REFERENCES}

\bibliographystyle{IEEEtran}
\bibliography{bibFile}

\begin{IEEEbiography}{Ashlesha Patil}{\,}completed her Ph.D. in Optics at the Wyant College of Optical Sciences, University of Arizona, in 2024, where her research focused on developing architectures and protocols for quantum networks. Prior to her doctoral studies, Ashlesha earned both her bachelor’s and master’s degrees in electrical engineering from IIT Bombay. She is currently working as a Quantum Architect at PsiQuantum.
\end{IEEEbiography}

\newpage

\begin{IEEEbiography}{Michele Pacenti}{\,}is a Ph.D. candidate at the University of Arizona. He received the B.S. and M.S. degrees (summa cum laude) in electronic engineering from the Polytechnic University of Marche, Italy, in 2019 and 2021 respectively. As a Center for Quantum Networks (CQN) affiliated student, he participated in the Convergent Quantum Research Alliance in
Telecommunications (CoQREATE) project, spending one month in 2023 as a visiting student at University College Dublin, Ireland. His interests include QLDPC code design, decoding algorithms, noise propagation in quantum circuits, and code-based entanglement distillation.
\end{IEEEbiography}

\begin{IEEEbiography}{Bane Vasi\'c}{\,}is a Professor of Electrical and Computer Engineering and Mathematics at the University of Arizona and a Director of the Error Correction Laboratory. Dr. Vasic leads the Quantum Error Correction Group within the Department of Energy multi-university project led by Fermi National Laboratory to establish a Center for Superconducting Materials and Systems. He is a co-PI of the NSF Center for Quantum Network hosted at the University of Arizona, and is a PI on seven research grants funded by the National Science Foundation. He is an IEEE Fellow, Fulbright Scholar, da Vinci Fellow, and a past Chair of IEEE Data Storage Technical Committee.
He is a founder of Codelucida, company developing advanced error correction solutions based on LDPC codes for solid state memories for data centers since 2012.  
\end{IEEEbiography}

\begin{IEEEbiography}{Saikat Guha}{\,} is the Clark Distinguished Chair Professor of Electrical and Computer Engineering at the University of Maryland, College Park. He received the B.Tech. degree in electrical engineering from the Indian Institute of Technology Kanpur and the S.M. and Ph.D. degrees from the Massachusetts Institute of Technology.
He was part of the founding team of the Quantum Information Processing Group at Raytheon BBN Technologies, where he worked from 2008 to 2017 as a Lead Scientist on photonic quantum information processing. From 2017 to 2024, he was with the University of Arizona, where he also held the Peyghambarian Endowed Chair in Optical Sciences and founded the NSF Center for Quantum Networks, which he continues to co-direct.
His research interests include information theory and quantum optics with applications to quantum sensing, communications, and networking. He was elected an IEEE Fellow in 2025 for contributions bridging information theory and physics in quantum-enhanced photonic systems.
\end{IEEEbiography}

\vspace{1cm}

\begin{IEEEbiography}{Narayanan Rengaswamy}{\,}is a tenure-track assistant professor in the Department of Electrical and Computer Engineering at the University of Arizona, Tucson, USA. His primary research is centered around quantum error correction and fault tolerance for quantum computing, communications, networking, and sensing. He is a Co-PI in the NSF Engineering Research Center for Quantum Networks (CQN) at the university. His work titled ``On the Optimality of CSS Codes for Transversal T'' was accepted as a Contributed Talk at the 2020 Quantum Information Processing conference (<25\% acceptance rate). He was a Keynote Speaker and Panelist at the 2024 Fault Tolerant Quantum Technologies workshop. He won Best Paper Awards in the Quantum Algorithms category at the 2024 and 2025 IEEE Quantum Week conferences for the papers titled ``Non-binary hypergraph product codes for qudit error correction'' and ``Fault Tolerant Quantum Simulation via Symplectic Transvections''. He was recognized as one of Computing’s Top 30 Early Career Researchers for 2025 by the IEEE Computer Society. He is an Editor for the journal ``Quantum'' and was one of two Lead Editors for the 2025 IEEE Journal on Selected Areas in Information Theory Special Issue on Quantum Error Correction and Fault Tolerance. He is a Senior Member of the IEEE and a Member of the AMS. 
\end{IEEEbiography}

\end{document}

%% file: sections/introduction.tex
\section{INTRODUCTION}
\label{sec:intro}
Shared entanglement over long distances enables entanglement-assisted sensing, distributed quantum computation, and quantum key distribution. Quantum networks distribute such entanglement using quantum repeaters~\cite{briegel1998quantum}, which store qubits in matter-based memories~\cite{guha2015rate} or photonic graph states~\cite{kaur2024resource} and perform local quantum operations.

Entanglement routing protocols generate Bell pairs between neighboring repeaters and connect them through joint measurements such as Bell-state measurements (BSMs)~\cite{pant2019routing} or Greenberger–Horne–Zeilinger (GHZ) projections~\cite{patil2020entanglement,patil2021distance}. We consider the realistic setting in which these Bell pairs are noisy and modeled as Werner states~\cite{victora2023entanglement}. Because Werner fidelity decays exponentially under successive BSMs, entanglement distillation is necessary to maintain usable long-distance entanglement.

Circuit-based distillation protocols~\cite{krastanov2019optimized,goodenough2023near} probabilistically consume multiple noisy pairs to produce fewer pairs of higher fidelity using local quantum circuits and classical communication. An alternative approach uses quantum error-correcting codes (QECCs)~\cite{kang2023trapped}, where stabilizer measurements across multiple noisy Bell pairs enable error correction and decoding into higher-fidelity pairs. Although circuit- and code-based distillation appear operationally distinct, they are equivalent in power~\cite{bennett1996mixed}. QECC-based distillation has been explored using convolutional codes~\cite{wilde_conv_dist} and extended to multipartite entanglement~\cite{rengaswamy_qldpc-ghz,rengaswamy2021distilling}.

In this work, we focus on two representative code families. Convolutional codes offer high rate and a streaming structure that enables continuous, low-latency distillation. Toric codes, by contrast, have low rate but strong error-correction capability, making them suitable for low-fidelity input states. These two extremes allow us to isolate the trade-off between rate and correction strength. Throughout, we assume depolarizing noise and Werner-state inputs.

Our contributions are threefold. First, we present a QECC-based distillation protocol and characterize its performance as a function of code rate and correction capability. Second, we introduce an entanglement-routing protocol for a chain of repeaters with probabilistic link generation and deterministic BSMs, in which a central processor selects where distillation is performed to maximize end-to-end distillable entanglement. Third, we provide a timing analysis that determines protocol latency and memory requirements.

The remainder of the paper reviews Werner states and QECCs, describes the distillation and routing protocols, evaluates performance for different codes, and concludes with future directions.

%% file: sections/background.tex
\section{BACKGROUND}
\label{sec:background}

\subsection{Werner states}

When the Bell state $\ket{\Phi^+}=\tfrac{\ket{00}+\ket{11}}{\sqrt{2}}$ undergoes depolarizing noise 
$\mathcal{N}:\rho \mapsto W\rho+\tfrac{1-W}{4}\mathbf{I}$, it becomes the Werner state
\begin{align}
    \rho_W =  W\ket{\Phi^+}\bra{\Phi^+} + \tfrac{1-W}{4}\mathbf{I},
    \label{eq:werner1}
\end{align}
where $W$ is the Werner parameter and $\mathbf{I}$ the identity. Expanding in the Bell basis gives
\begin{align}
\rho_W = F\ket{\Phi^+}\bra{\Phi^+}
        + \frac{1-F}{3}\!\!\sum_{\beta\in\{\Phi^-,\Psi^+,\Psi^-\}}\!\!\ket{\beta}\bra{\beta}.
\end{align}

with fidelity $F=\tfrac{3W+1}{4}$. Depolarizing noise is information-theoretically the hardest to correct, making it a natural benchmark for distillation protocols.  

A Bell-state measurement (BSM) is a joint projective measurement on two qubits that projects them onto one of the four Bell states, $\ket{\Phi^\pm}$ or $\ket{\Psi^\pm}$. It is a key primitive for entanglement swapping: when two parties each share an entangled pair and the intermediate qubits are measured in the Bell basis, the measurement outcome entangles the remaining, previously uncorrelated qubits, up to known Pauli corrections. This allows distant parties to establish entanglement without their qubits ever interacting directly.

A BSM on Werner states with fidelities $F_1,F_2$ produces another Werner state of fidelity~\cite{victora2023entanglement}
\begin{align}
F = \tfrac{1}{4} + \tfrac{3}{4}\left(\tfrac{4F_1 - 1}{3}\right)\left(\tfrac{4F_2 - 1}{3}\right)
\end{align}
or, equivalently, $W=W_1W_2$. Since $F<F_1,F_2$ for $F_1,F_2<1$, fidelities decay along a chain of $n$ repeaters as  
\begin{align}
F = \tfrac{1}{4} + \tfrac{3}{4}\prod_{i=1}^{n+1}\left(\tfrac{4F_i - 1}{3}\right),
\label{eq:BSM_werner}
\end{align}
where $F_i$ is the fidelity of the $i$-th link.  

The distillable entanglement of a Werner state is defined as~\cite{bennett1996mixed}:
\begin{equation}
\small
D(F) = 
  \begin{cases}
1+F\log_2F+(1-F)\log_2\!\tfrac{1-F}{3} & \mathrm{if}\quad F>0.8107 \\
 0 & \mathrm{otherwise}.
\end{cases}  
\label{eq:dist_ent}
\end{equation}  

\subsection{Quantum error correction for entanglement distillation}

\subsubsection{Stabilizer formalism}
Let $I,X,Y,Z$ be the Pauli operators and $\mathcal{G}_n$ the $n$-qubit Pauli group, i.e., tensor products of these operators with phases $\{\pm1,\pm i\}$. A stabilizer group $\mathcal{S}$ is an Abelian subgroup of $\mathcal{G}_n$ not containing $-I$, generated by $n-k$ independent elements. If $k=0$, $\mathcal{S}$ defines a unique \emph{stabilizer state}.  

Pauli measurements update $\mathcal{S}$, allowing the \emph{stabilizer formalism}~\cite{patil2023clifford} to track the post-measurement state. For example, the $n$-fold Bell state $\ket{\Phi^+_n}_{AB}$ is stabilized by $\langle X_AX_B,Z_AZ_B\rangle^{\otimes n}$ and satisfies the \emph{Bell state matrix identity}~\cite{bennett1996mixed}:
\begin{equation}
(M_A \otimes I_B)\ket{\Phi^+_n}_{AB} = (I_A \otimes M_B^T)\ket{\Phi^+_n}_{AB},
\label{eq:bell_state_id}
\end{equation}
for any $n$-qubit operator $M$. If $M$ is a projector, i.e., $M^2 = M$, then one can replace $M$ by $M^2$ and apply this property to only one $M$ so that the operator on the right-hand side becomes $(M_A \otimes M_B^T)$. This means that projecting one qubit of the Bell pair by $M$ induces a simultaneous projection of the other qubit by $M^T$.

\subsubsection{Stabilizer codes}
An $\llbracket n,k,d \rrbracket$ stabilizer code encodes $k$ qubits into $n$ as the joint $+1$ eigenspace of $n-k$ generators of the Pauli group. The minimum distance $d$ is the smallest weight of an operator in the normalizer $\mathcal{N}(\mathcal{S})$ but not in $\mathcal{S}$. Measuring the stabilizers yields a \emph{syndrome} $\mathbf{s}$, which is trivial if the error lies in $\mathcal{N}(\mathcal{S})$ and non-trivial otherwise. A decoder uses $\mathbf{s}$ to estimate the error $\hat{E}$; recovery succeeds if $\hat{E}E\in\mathcal{S}$ and fails with a \emph{logical error} otherwise.

\subsubsection{The Toric code}
It is a $\llbracket 2d^2,2,d \rrbracket$ stabilizer code that can be described by a $d\times d$ square lattice whose opposite boundaries are identified with each other (to make a torus). Here, qubits reside on edges and the set of stabilizers is given by the incidence relations between vertices, edges and faces~\cite{dennis_topological_2002}. Specifically, each face or plaquette (resp. vertex) represents a weight-$4$ $Z$-type (resp. $X$-type) stabilizer acting on the four incident edges. It can be shown that the minimum distance of the code is indeed the lattice size, $d$. For decoding the toric code we consider the minimum-weight perfect-matching (MWPM) decoder~\cite{qecsim}. 

\subsubsection{Quantum convolutional codes}
A quantum convolutional code (QCC) is defined by $n-k$ commuting Pauli sequences $S_0 = \{s_j, 1 \leq j \leq n-k\}$, where $s_j$ is of length $(\nu+1) n$ and $\nu$ is called the \textit{memory} of the code. The full stabilizer $\mathbf{S}$ is infinite, which contains $S_0$ as well as the shifts of $S_0$ by multiples of $n$ qubits, denoted as $S_i,\ i \in \mathbb{N}$.

In this paper we consider the $\llbracket 3,1,3 \rrbracket$ QCC over $\mathbb{F}_4$ from \cite{forney2007convolutional}. Its stabilizer matrix is the generator matrix of a classical rate-$1/3$ quaternary self-orthogonal convolutional code with memory $\nu = 1$, which is generated by the following generator as well as all its shifts by multiples of $n=3$ symbols:
\begin{equation}
s_1 = 
\begin{bmatrix}
1 & 1 & 1 & 1 & \omega & \bar{\omega}
\end{bmatrix}.
\end{equation}
Note that $\bar{\omega} = \omega^2$ in $\mathbb{F}_4$.
To perform the decoding, we use a quaternary syndrome Viterbi decoder~\cite{syndromeviterbi}. The Viterbi decoder has the advantage to act as a \textit{window} decoder, that can correct errors in an online fashion with maximum likelihood performance, without having to wait for the whole Bell pair sequence to arrive at the receiver.

\subsubsection{Code-based entanglement distillation}
\label{subsubsub:QECC_dist}
QECCs can be used for entanglement distillation~\cite{rengaswamy_qldpc-ghz}. Suppose Alice and Bob want to share high-fidelity Bell pairs, and a central repeater Charlie generates $n$ Werner states with parameter $W$. The $2n$ qubits are described by the stabilizer group $\mathcal{S}_W(\ket{\Phi^+_n}_{AB})$. Label the halves sent to Alice and Bob as $A$ and $B$, respectively. Charlie projects his $A$ qubits onto the code space of an $\llbracket n,k,d\rrbracket$ stabilizer code by measuring its $n-k$ stabilizers, updating the stabilizer group via the stabilizer formalism. By the Bell state identity~\eqref{eq:bell_state_id}, Bob’s halves are simultaneously projected onto the transpose code, yielding $2(n-k)$ stabilizers and $2k$ logical operators, which correspond to $k$ encoded Bell pairs shared between Alice and Bob. The outcome of this measurement is a random syndrome, which defines the code space along with the stabilizers of the code.

Charlie then sends the qubits $A$ and $B$ along with the measured syndrome. They measure all stabilizers again to obtain an error syndrome, and use a decoder to correct errors. If the code corrects the noise and decoding succeeds, the result is $k$ perfect Bell pairs encoded in two $\llbracket n,k,d\rrbracket$ codes. If a logical error occurs, some pairs flip into orthogonal Bell states, and both parties must succeed for distillation to work. This procedure also applies to quantum convolutional codes~\cite{wilde_conv_dist}. Finally, Alice and Bob recover the $k$ distilled Bell pairs by applying the inverse encoding to their $n$ physical qubits. 

\begin{table}
\begin{center}
\begin{tabularx}{\columnwidth}{lX}
\hline
\textbf{Symbol} & \textbf{Description} \\
\hline
\multicolumn{2}{l}{\textit{Network and Topology}} \\
$N$ & Number of repeaters in the chain \\
$L$ & Total distance between Alice and Bob \\
$\ell$ & Distance between adjacent nodes ($L/(N+1)$) \\
$M$ & Spatial multiplexing parameter (modes per slot) \\
$p$ & Link (BSM) success probability \\
$c_t$ & Speed of light in optical fiber \\
\hline
\multicolumn{2}{l}{\textit{Quantum States and QECC}} \\
$\rho_W$ & Werner state \\
$W$ & Werner parameter \\
$F, F_0$ & State fidelity (general) and initial link fidelity \\
$D(F)$ & Distillable entanglement of a state with fidelity $F$ \\
$[n, k, d]$ & QECC parameters: physical qubits, logical qubits, distance \\
$R$ & End-to-end entanglement rate \\
\hline
\multicolumn{2}{l}{\textit{Protocol \& Resource Management}} \\
$\pi$ & Scheduling policy (composition of distillation/BSM nodes) \\
$Q_{\mathrm{BSM}}$ & Memory usage (qubits) at a BSM repeater \\
$Q_{\mathrm{Dist}}$ & Memory usage (qubits) at a distillation repeater \\
\hline
\multicolumn{2}{l}{\textit{Timing}} \\
$\tau$ & Time slot duration \\
$\tau_{\mathrm{BSM}}$ & Time required to perform a Bell State Measurement \\
$\tau_{\mathrm{Dec}}$ & Classical decoding time for the QECC \\
$\tau_{\mathrm{CC}}$ & Time to transmit BSM outcomes to Bob \\
\hline
\end{tabularx}
\end{center}
\caption{Summary of notation.}
\label{tab:notation}
\end{table}

%% file: sections/protocol.tex
\section{THE GLOBAL LINK-STATE PROTOCOL}
\label{sec:protocol}
We present an entanglement routing protocol assuming quantum memories with coherence times longer than all network timescales, so stored Werner states do not degrade. We first outline QECC-based distillation, then describe the routing protocol and its latency. In Table \ref{tab:notation} we summarize all the notational symbols we utilize throughout the paper.

\subsection{Entanglement distillation}
\label{subsec:distillation}

We use $\llbracket n,k,d\rrbracket$ QECCs to distill $n$ Werner states into $k$ higher-fidelity states, and now compute their output fidelity and success probability.  
\begin{figure}
    \centering
\includegraphics[width=\linewidth]{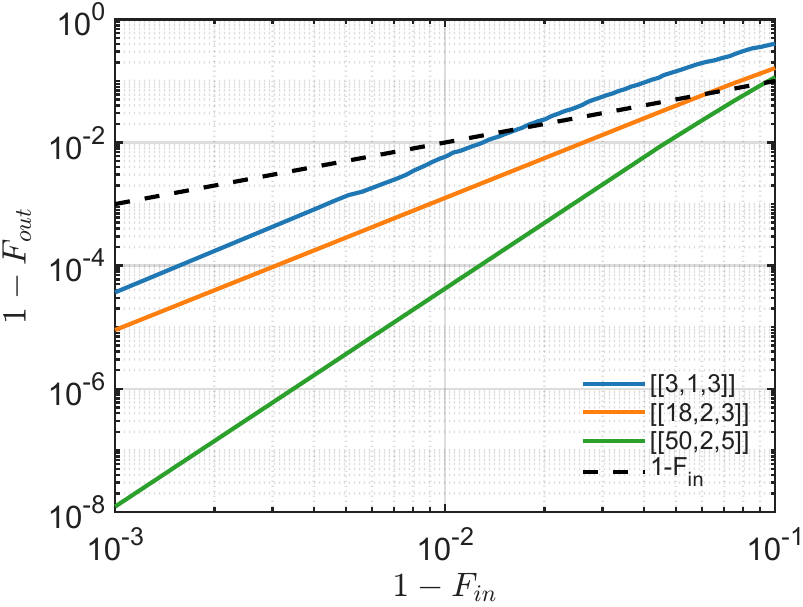}
    \caption{Output infidelity $(1-F_{\rm out})$ vs. input infidelity $(1-F_{\rm in})$ for distillation with different QECCs. The $\llbracket 3,1,3 \rrbracket$ code is convolutional; the others are toric.}
    \label{fig:distillation}
\end{figure}
Toric and convolutional codes are decoded with MWPM and Viterbi algorithms respectively, which always converge to a valid error estimate, yielding a distillation success probability of 1. Logical errors, however, lead to residual infidelity in the output states. We therefore map input infidelity $(1-F_{\rm in})$ to output infidelity $(1-F_{\rm out})$ using the logical error rate of the QECC, as shown in Figure~\ref{fig:distillation}. We assume perfect syndrome measurements; more realistic models would only reduce the output fidelity. Each code has a break-even point $F_{\rm th}$ such that for $F_{\rm in}<F_{\rm th}$ distillation worsens, rather than improves, fidelity.

\subsection{The protocol}
\label{subsec:protocol1}
We consider a linear chain of $N$ repeaters connecting Alice and Bob over distance $L$, assuming perfect quantum memories. Between each pair of repeaters, an \emph{intermediate node} performs photonic BSMs. In each time slot of duration $\tau$, every repeater generates $2M$ Werner states and sends $M$ qubits through parallel optical channels toward each neighboring intermediate node, where $M$ is the spatial multiplexing parameter. Successful BSMs (with probability $p$) create elementary links of fidelity $F_0$.
The resulting network configuration is called a \emph{snapshot}, and the collection of all snapshots constitutes the global link-state information. Each intermediate node reports its local snapshot to a central processor, which determines an optimal \emph{scheduling strategy}, that is, it selects which repeaters act as distillation repeaters and which perform only BSMs. This decision procedure is described in the ``Scheduling strategy'' subsection and is then communicated classically to the repeaters.
BSM repeaters first perform BSMs, producing longer-range links between distillation repeaters, generally with fidelity below $F_0$. At distillation repeaters, available links are grouped into blocks of $n$ and distilled using an $\llbracket n,k,d\rrbracket$ QECC, converting $n$ noisy links into $k$ higher-fidelity ones, while leftover links remain undistilled. Distillation is applied only when it increases fidelity, as determined from~\eqref{eq:BSM_werner} and Fig.~\ref{fig:distillation}. Finally, each repeater performs BSMs on its remaining links, yielding end-to-end entanglement between Alice and Bob.

\begin{figure}[t]
    \centering
    \includegraphics[width=1\linewidth]{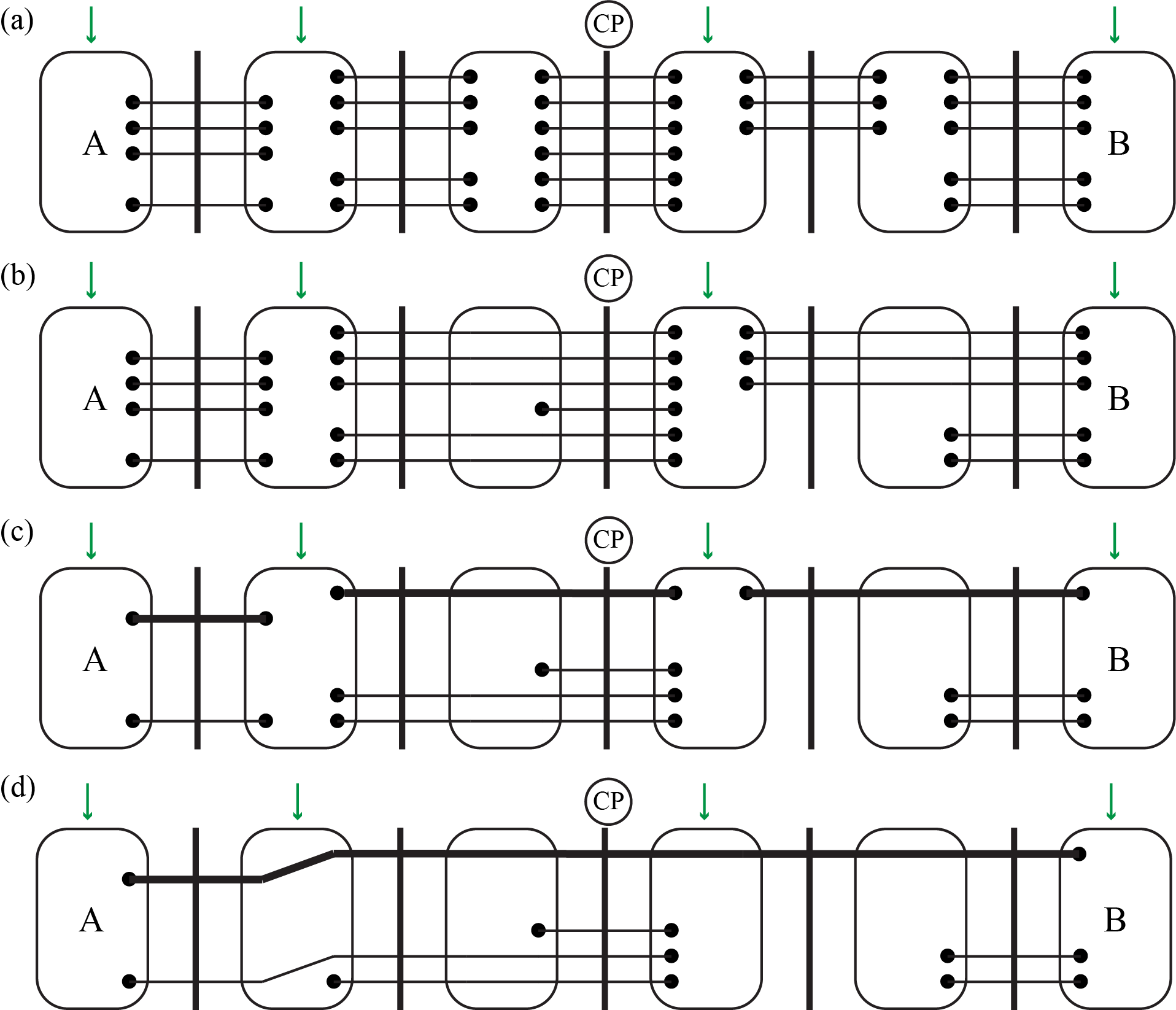}
    \caption{Schematic of the protocol with $N=4$, $M=6$, $n=3$, and $k=1$. The central processor is located halfway between Alice and Bob (circle). (a) Network snapshot showing links (black lines) between neighboring quantum memories (black circles). Vertical lines represent intermediate nodes. Distillation repeaters are marked with green arrows; the others are BSM repeaters. (b) BSM repeaters perform BSMs on all available links. (c) Distillation repeaters group the links into triples and distill one higher-fidelity link (thick black line). (d) Distillation repeaters, except for Alice and Bob, perform BSMs to create end-to-end links.}
    \label{fig:schematic}
\end{figure}
A schematic of the protocol is illustrated in Figure~\ref{fig:schematic}.
\begin{figure}
    \centering
    \input{figs/tikz_figures/latency.tikz}
    \caption{Timing diagram to illustrate the latency of the protocol. Network nodes include Alice, Bob, and the repeaters. Here, $\tau_{\rm BSM}<\tau_{\rm Dist}<(\tau_{\rm BSM}+\tau_{\rm CC})$.}
    \label{fig:latency}
\end{figure}

\subsection{Scheduling strategy}
\label{sub:brute-force}
The central processor determines the optimal sequence of BSM and distillation operations via an exhaustive search over all compositions of $(N+1)$. A composition $\pi=(n_1,\dots,n_i)$ is an ordered partition of an integer, and specifies the hop distances between distillation repeaters, with the remaining nodes acting as BSM repeaters. For example, with $N=2$ links, $\pi=(1,1,1)$ corresponds to distillation at every node, while $\pi=(3,0)$ corresponds to no distillation. Notice that $N$ includes the end nodes Alice and Bob.
For each $\pi$, the processor evaluates the resulting end-to-end entanglement rate. If the final links have fidelities $F_1^\pi,\dots,F_M^\pi$, the rate is defined as
\begin{equation}
\label{eq:dist_rate}
R(\pi)=\frac{\sum_{i=1}^M D(F_i^\pi)}{2\tau M}
=\frac{D_{\mathrm{total}}(\pi)}{2\tau M}.
\end{equation}
Although the same QECC is used for all blocks, the fidelities $F_i^\pi$ may differ because different links undergo different numbers of BSMs before distillation.
The optimal scheduling is then
\begin{equation}
\pi_{\mathrm{opt}}=\arg\max_{\pi} D_{\mathrm{total}}(\pi).
\end{equation}
We assume uniform repeater spacing for simplicity; non-uniform distances can be handled within the same fidelity-based scheduling framework. All computations are classical and performed by the central processor, while repeaters execute only the prescribed quantum operations.

\begin{figure*}
    \centering
    \subfloat[$F_0 = 0.99$]{%
        \includegraphics[width=0.48\textwidth]{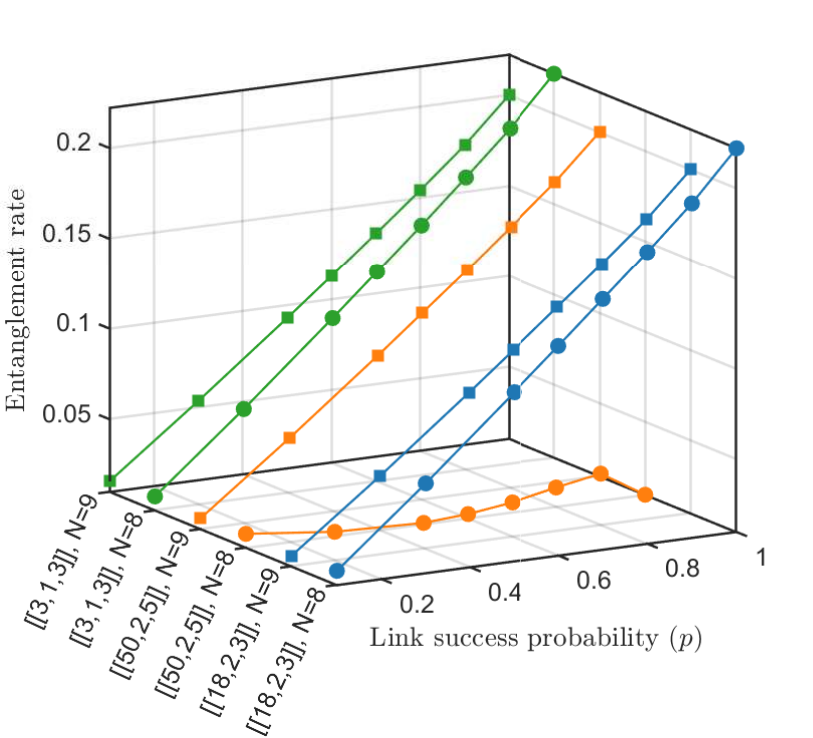}%
        \label{fig:compare1850_099}
    }\hfill
    \subfloat[$F_0 = 0.97$]{%
        \includegraphics[width=0.48\textwidth]{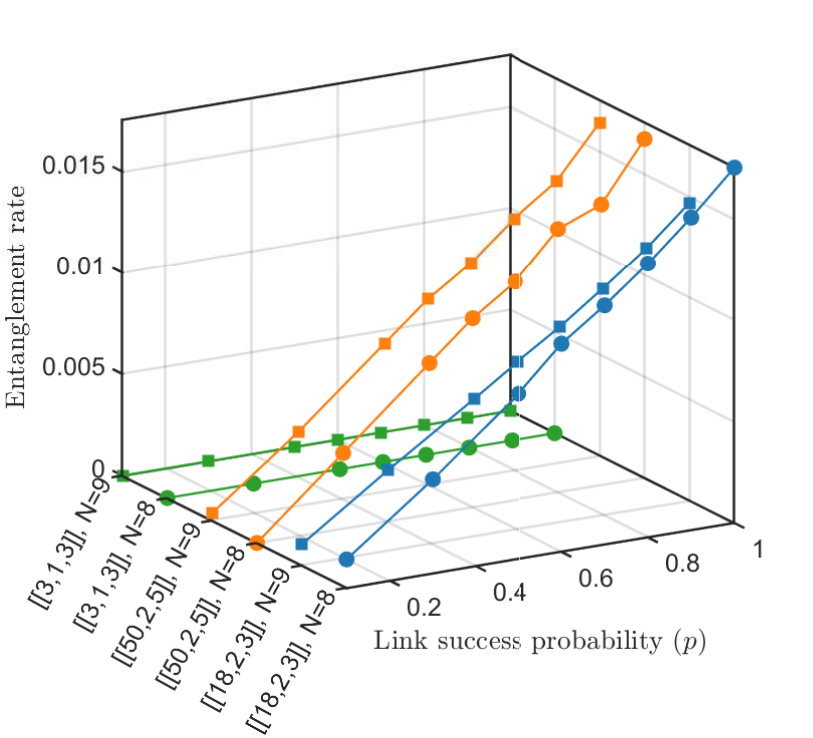}%
        \label{fig:compare1850_097}
    }
    \caption{Distillable entanglement as a function of $p$ when the $\llbracket 18,2,3\rrbracket$ and $\llbracket 50,2,5\rrbracket$ surface codes and the $\llbracket 3,1,3\rrbracket$ convolutional code are used for distillation, with $M=450$.}
    \label{fig:compare1850_merged}
\end{figure*}

\subsection{Latency of the protocol}

The repetition rate of the protocol is $1/\tau$. The distance between each pair of repeaters is equal to $\ell = L/(N+1)$, and since each BSM processor is placed in between each link, the distance traveled by photons and classical information is $\ell/2$, which corresponds to a travel time of:
\begin{equation}
\tau_l = \frac{L}{2(N+1)c_f},
\end{equation}
where $c_f$ is the speed of light in fiber. Then, each BSM node must send its snapshot to the central node, which for simplicity is assumed to be placed exactly in the middle between Alice and Bob. For the farthest BSM node, the distance with the central processor is $\ell \times N/2$; then, the central processor communicates the scheduling strategy to all the repeaters, the farthest being located at distance $L/2$. Therefore, the total processing latency is given by:
\begin{equation}
\tau_p = \frac{1}{c_f}\left(\frac{\ell N}{2}+\frac{L}{2}\right) =  \frac{(2N+1)L}{2(N+1)c_f}.
\end{equation}
The decoding time $\tau_{\rm Dec}=10\tau$ typically depends on the specific decoder chosen and by the code length, although here we simply assume it to be large enough to capture the effect introduced by its latency time. Once the distillation is complete, every repeater communicates the Pauli corrections from distillation to the central processor, which then transmits all the corrections to Bob, with a decoding time of:
\begin{equation}
\tau_{\rm Dist} = \frac{L}{c_f}.
\end{equation}
All repeaters send the BSM outcomes to Bob. The upper limit for the time required to send the BSM outcomes is the time it takes to send them from the repeater farthest from Bob:
\begin{equation}
\tau_{\rm CC} = \frac{NL}{(N+1)c_f}.
\end{equation}
Furthermore, we call the time required to perform a BSM $\tau_{BSM}$. Assuming $\tau_{\rm BSM} < \tau_{\rm Dist} < (\tau_{\rm BSM} + \tau_{\rm CC})$, Bob applies the Pauli corrections for the first end-to-end entangled state at time
\begin{equation}
    \tau = \tau_l + \tau_p + 2\tau_{\rm BSM} + \tau_{\rm Dec} + \tau_{\rm CC},
\end{equation}
which is the total classical communication latency of the protocol, as shown in Figure~\ref{fig:latency}. The correction time $\tau_{\rm Dist}$ does not affect the latency, since the Pauli corrections from distillation reach Bob before those from the BSM outcomes.

%% file: figs/tikz_figures/latency.tikz
\begin{center}
\resizebox{1\linewidth}{!}{
\begin{tikzpicture}[
  font=\sffamily\footnotesize,
  >=Stealth,thick,
  commentl/.style={text width=4.0cm, align=right,font=\sffamily\scriptsize}, 
  commentr/.style={commentl, align=left}
]

\node[] (init) {Nodes};
\node[right=2cm of init] (recv) {Processor};

\draw[->] ([yshift=-1.7cm]init.south) coordinate (fin1o) -- 
          ([yshift=-.7cm]fin1o-|recv) coordinate (fin1e) 
          node[pos=.5, above, sloped, font=\tiny] {Snapshot};

\draw[->] ([yshift=0cm]fin1e) coordinate (ack1o) -- 
          ([yshift=-.7cm]ack1o-|init) coordinate (ack1e) 
          node[pos=.5, below, sloped, font=\tiny] {Scheduling strategy};

\draw[->] ([yshift=-4.8cm]fin1e) coordinate (fin2o1) -- 
          ([yshift=-.99cm]fin2o1-|init) coordinate (fin2e1) 
          node[pos=.46, below, sloped, font=\tiny] {Distillation\ \ \  corrections};

\draw[->] ([yshift=-4.3cm]fin1o) coordinate (ack2o1) -- 
          ([yshift=-1.2cm]ack2o1-|recv) coordinate (ack2e1) 
          node[pos=.5, above, sloped, font=\tiny] {Distillation corrections};

\draw[->] ([yshift=-6cm]fin1e) coordinate (fin2o2) -- 
          ([yshift=-.99cm]fin2o2-|init) coordinate (fin2e2) 
          node[pos=.5, above, sloped, font=\tiny] {BSM corrections};

\draw[->] ([yshift=-5.5cm]fin1o) coordinate (ack2o) -- 
          ([yshift=-1.2cm]ack2o-|recv) coordinate (ack2e) 
          node[pos=0.25, above, sloped, font=\tiny] {BSM corrections};

\draw[thick, shorten >=-1cm] (init) -- (init|-ack2e);
\draw[thick, shorten >=-1cm] (recv) -- (recv|-ack2e);

\node[left = 2mm of fin1o.west, commentl]{{\itshape Links created.}\\[-1mm]{$\tau_l$}};
\node[left = 2mm of ack1e.west, commentl]{\itshape Strategy received.\\ \itshape BSMs start.\\[-1mm]{$\tau_l+\tau_p$}};
\node[below left = 6mm and 2mm of ack1e.west, commentl]{\itshape BSMs complete.\\ \itshape Distillations start.\\[-1mm]{$\tau_l+\tau_p+\tau_{\rm BSM}$}};
\node[below left = 19mm and 2mm of ack1e.west, commentl]{\itshape Distillation complete.\\ \itshape BSMs at distillation\\\itshape repeaters start.\\[-1mm]{$\tau_l+\tau_p+\tau_{\rm BSM}+\tau_{\rm Dec}$}};
\node[below left = 35mm and 2mm of ack1e.west, commentl]{\itshape BSMs complete.\\ \itshape End-to-end entanglement.\\[-1mm]{$\tau_l+\tau_p+2\tau_{\rm BSM}+\tau_{\rm Dec}$}};
\node[below left = 49mm and 2mm of ack1e.west, commentl]{\itshape Received by Bob.\\[-1mm]{$\tau_l+\tau_p+\tau_{\rm BSM}+\tau_{\rm Dec}+\tau_{\rm Dist}$}};
\node[below left = 58mm and 2mm of ack1e.west, commentl]{\itshape Received by Bob.\\[-1mm]{$\tau_l+\tau_p+2\tau_{\rm BSM}+\tau_{\rm Dec}+\tau_{\rm CC}$}};

\draw[->,thick] ([xshift=-3.35cm]init.west) -- ([xshift=-3.9cm] fin2e2)
  node[midway,above,font=\sffamily\scriptsize,rotate=90] {Time};

\end{tikzpicture}
}
\end{center}

%% file: sections/results.tex
\section{RESULTS}
\label{sec:results}

\subsection{Rate calculations}

We compare the $\llbracket 3,1,3 \rrbracket$ quantum convolutional code with the $\llbracket 18,2,3 \rrbracket$ and $\llbracket 50,2,5 \rrbracket$ toric codes under the global link-state protocol. A protocol ${n,N}$ employs an $\llbracket n,k,d \rrbracket$ code across $N$ repeaters. We assume perfect quantum memories and set $F_0={0.99,0.97}$ to probe the above- and below-threshold regimes. We choose $M=450$ to be large enough compared to the length of the code. Figure~\ref{fig:compare1850_099} reports the distillable entanglement rate for $F_0=0.99$.

The entanglement rate generally increases with the link success probability $p$, except for the ${50,8}$ protocol. As $p\to1$, ${50,8}$ converges to the $(1,1,\dots,1)$ composition, performing distillation between all adjacent repeaters. Due to the low rate of the $\llbracket 50,2,5 \rrbracket$ code, this yields only $M k/n = 18$ high-fidelity end-to-end Bell pairs, with fidelity approaching unity, as shown in Table~\ref{tab:compare_merged}.

By contrast, protocols such as ${50,9}$, ${3,8}$, and ${18,8}$ converge to the $(N+1,0)$ composition, corresponding to BSM-only operation. In these cases, the fidelity of long-range links falls below the break-even point of the code, so distillation is ineffective. For ${50,9}$, the additional repeater introduces an extra BSM that reduces the input fidelity just enough to prevent distillation. Similar behavior occurs for the higher-rate $\llbracket 3,1,3 \rrbracket$ and $\llbracket 18,2,3 \rrbracket$ codes, which are effective only for near-ideal inputs.

The trade-off between break-even fidelity and code rate also explains why ${50,8}$ achieves a higher rate at $p=0.9$ than at $p=1$. For $p<1$, only a fraction $pM$ of links are generated, and leftover links that cannot be grouped into blocks of 50 undergo BSM-only connections. Although noisier, their larger number increases the total distillable entanglement rate.

For $F_0=0.97$ (Fig.~\ref{fig:compare1850_097}), the $\llbracket 3,1,3 \rrbracket$ code lies below threshold and yields zero rate, while the ${18,8}$, ${18,9}$, ${50,8}$, and ${50,9}$ protocols remain above threshold and select the $(1,1,\dots,1)$ composition, with fidelity decreasing as $N$ increases.

Overall, for large $p$, the protocol selects between $(1,1,\dots,1)$ and $(N+1,0)$ depending on whether the long-range link fidelity exceeds the code’s break-even point. High-rate codes require near-ideal inputs, whereas low-rate codes operate at lower fidelities but yield fewer final states.

\begin{table}[t]
\centering
\setlength{\tabcolsep}{3pt}
\begin{tabular}{c c cc cc cc}
\toprule
 &  & \multicolumn{2}{c}{$\llbracket 3,1,3\rrbracket$} &
        \multicolumn{2}{c}{$\llbracket 18,2,3\rrbracket$} &
        \multicolumn{2}{c}{$\llbracket 50,2,5\rrbracket$} \\
\cmidrule(lr){3-4} \cmidrule(lr){5-6} \cmidrule(lr){7-8}
& $N$ & 8 & 9 & 8 & 9 & 8 & 9 \\
\midrule
\multirow{2}{*}{$F_0 = 0.99$}
 & Avg. fidelity & 0.91 & 0.90 & 0.91 & 0.90 & 0.99 & 0.90 \\
 & E2E states    & 450  & 450  & 450  & 450  & 18   & 450 \\
\midrule
\multirow{2}{*}{$F_0 = 0.97$}
 & Avg. fidelity & 0.77 & 0.75 & 0.89 & 0.88 & 0.98 & 0.98 \\
 & E2E states    & 450  & 450  & 50   & 50   & 18   & 18 \\
\bottomrule
\end{tabular}
\vspace{5pt}
\caption{Number of end-to-end states and average fidelity for different QECCs when $p=1$ and $M=450$, for two values of the elementary-link fidelity $F_0$.}
\label{tab:compare_merged}
\end{table}

\subsection{Quantum memory requirements}

\begin{figure}
    \centering
    \includegraphics[width=1\linewidth]{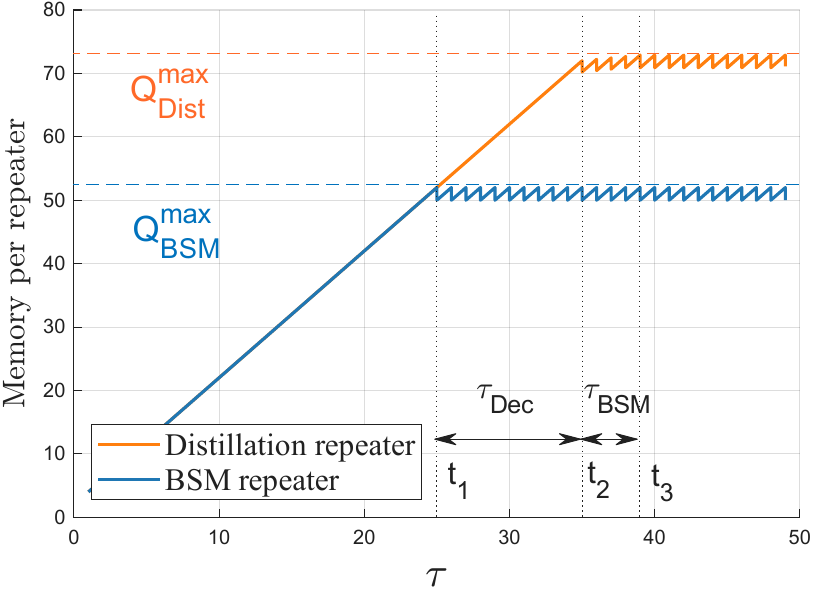}
    \caption{The number of quantum memories used at distillation and BSM repeaters for the $\llbracket 50,2,5\rrbracket$ code. Here, $\tau_p=20\tau, \tau_l=\tau, \tau_{\rm BSM}=4\tau, \tau_{\rm Dec}=10\tau$. }
    \label{fig:memories}
\end{figure}

In this section, we derive an upper bound on the number of quantum memories required per repeater. Memory usage depends on the network timescales and on the rate of the error-correcting code.

\begin{lemma}
Let $M$ be the number of parallel links initiated every time step $\tau$. Let $Q_{\mathrm{BSM}}(t)$ and $Q_{\mathrm{Dist}}(t)$ denote the number of quantum memories in use at time $t$ for a Bell-state-measurement (BSM) repeater and a distillation repeater, respectively. Assume $\tau_l$, $\tau_p$, $\tau_{\rm BSM}$, and $\tau_{\rm Dec}$ are integer multiples of $\tau$. Then the maximum memory occupancies satisfy
\begin{equation}
\label{eq:latency1}
Q_{\mathrm{BSM}}^{\max}
= 2M\left(\frac{\tau_l + \tau_p + \tau_{\rm BSM}}{\tau} + 2\right),
\end{equation}
and
\begin{equation}
\label{eq:latency2}
\begin{aligned}
Q_{\mathrm{Dist}}^{\max}
= {} & 2M\Bigg[
\frac{\tau_l + \tau_p + 2\tau_{\rm BSM} + \tau_{\rm Dec}}{\tau}
\\
& \quad + 1
- \left(1 + \frac{\tau_{\rm BSM}}{\tau}\right)\frac{n-k}{n}
\Bigg].
\end{aligned}
\end{equation}
where $(n,k)$ are the QECC parameters.
\end{lemma}

\begin{proof}
At each time step $\tau$, the repeater allocates $2M$ new memories for link generation, so prior to any release events the memory usage grows as $2M(t/\tau+1)$.

For a BSM repeater, the first release occurs at $t_1=\tau_l+\tau_p+\tau_{\rm BSM}$, when the earliest links are measured and freed. The worst-case occupancy occurs immediately before a subsequent release, after a new allocation but before memory is freed, yielding (\ref{eq:latency1}).

For a distillation repeater, memory accumulates until distillation at
$t_2=\tau_l+\tau_p+\tau_{\rm BSM}+\tau_{\rm Dec}$. An $(n,k)$ distillation reduces the number of stored links by a factor $(n-k)/n$. The peak occupancy occurs just before BSMs are applied to the distilled links at
$t_3=\tau_l+\tau_p+2\tau_{\rm BSM}+\tau_{\rm Dec}$, giving (\ref{eq:latency2}).
\end{proof}

Figure~\ref{fig:memories} illustrates the memory occupancy over time for $p=1$, highlighting the critical times $t_1$, $t_2$, and $t_3$ for the $\llbracket 50,2,5 \rrbracket$ toric code. Distillation repeaters generally require more memory than BSM repeaters due to the decoding delay $\tau_{\rm Dec}$. When $\tau_{\rm Dec}\ll\tau$, the memory requirements approach those of BSM repeaters. Higher-rate codes further increase memory usage by producing more distilled links per block.

\section{CONCLUSION AND DISCUSSION}
\label{sec:conclusion}
We proposed a protocol for routing Werner states across a repeater chain using QECC-based distillation and link-state information to select where distillation is performed. A scheduling strategy maximizes end-to-end distillable entanglement for each network snapshot. Code rate governs the trade-off between fidelity and yield: low-rate codes improve fidelity at the cost of consuming more links, while high-rate codes increase throughput at lower fidelity. Multiplexing partially compensates the throughput loss of low-rate codes.

Code rate also determines memory usage. For an $\llbracket n,k,d\rrbracket$ QECC, the fraction of freed memories is $(n-k)/n$, so lower-rate codes reduce memory requirements assuming sufficiently fast decoding. Extensions to finite-coherence memories follow by incorporating decoherence during communication and processing.

With uniform link success probabilities, the optimal composition is one of two extremes, depending on $N$, $F_0$, and the code rate. Non-uniform link success probabilities or heterogeneous code choices across repeaters may favor intermediate compositions~\cite{cheng2025adaptiveerrorcorrectionentanglement}.

Although the protocol assumes global link-state knowledge, the same decisions can be made locally. A repeater using only its position, $F_0$, and the QECC can decide whether to perform distillation, reducing latency to $2\tau_l+2\tau_{\rm BSM}+\tau_{\rm Dec}+\tau_{\rm CC}$ and lowering memory requirements by $2M(\tau_p-\tau_l)/\tau$.

\section{ACKNOWLEDGMENTS}
This work was co-funded by the National Science Foundation (NSF) ERC Center for Quantum Networks under grant number 1941583 and NSF grant number CIF-2106189. The work of Michele Pacenti and Bane Vasi\'c is also supported by NSF under grants CIF-1855879, CCF-2100013 and ECCS/CCSS-2027844, and by a generous gift from Maecenases Dora and Barry Bursey. Bane Vasi\'{c} has disclosed an outside
interest in his startup company Codelucida to The University
of Arizona. Conflicts of interest resulting from this interest are
being managed by The University of Arizona in accordance
with its policies.